\documentstyle[amssymb]{article}
 \newcommand {\nc}{\newcommand}
 \nc{\nn}{\nonumber}
 \nc{\eq}{\begin{equation}}
 \nc{\en}{\end{equation}}
 \nc{\eqa}{\begin{eqnarray}}
 \nc{\ena}{\end{eqnarray}}
 \nc {\norm}[1]{\parallel{#1}\parallel}
 \def\intg{{\mathbb Z}}
 \def\real{{\mathbb R}}
 \def\complex{{\mathbb C}}
 \def\Alg{{\mathcal A}}

 \def\lag{{\cal L}}

 \def\pr{\prime}
 \def\ot{\otimes}
 \def\dt{\delta}
 \def\pt{\partial}
 \def\dts{\dt_\sigma}
 
 \def\etal{{\it et al} }
 \newtheorem{definition}{Definition}
 \newtheorem{lemma}{Lemma}
 \newtheorem{theorem}{Theorem}
 \newtheorem{proposition}{Proposition}

 \newtheorem{remark}{Remark}
 
 \def\prf{{\bf{Proof:}}\\}
 \def\endprf{${\bf{\Box}}$\\}
 \def\CDalign#1{\bgroup\vcenter\bgroup\tabskip 2pt 
 \baselineskip 14pt \lineskip 3pt \lineskiplimit 3pt
 \halign\bgroup &\hfill$##$\hfill\crcr
 #1\crcr\egroup\egroup\egroup} 
 \def\CDdown{\Big\downarrow}       
 \def\CDrlabel#1{\vcenter{\hbox to0pt{$\scriptstyle#1$\hss}}} 
 \def\CDto{\mathop{\relbar\joinrel\longrightarrow}\limits}    

 \title{Noncommutative Differential Geometry and Classical Field Theory on Finite Groups
 \footnote{Based on the report on the Seminar ``Field Theory, Gravity and Symplectic Structure''
 [China Center of Advanced Science and Technology (CCAST) \& World Laboratory (WL), Beijing, July 9th to July 13rd,
 2001]}
 }
 \author{Jian Dai, Xing-Chang Song\\
  Institute of Theoretical Physics, School of Physics, Peking University \\
  Beijing, P. R. China, 100871\\
  jdai@mail.phy.pku.edu.cn, songxc@ibm320h.phy.pku.edu.cn
  }
 \date{July 10th, 2001}
 
\begin{document}
  \maketitle
  \begin{abstract}
  \noindent
  Plan of this report
  is given below
  \begin{enumerate}
   \item Motivation from Physical and Mathematical Point of View
   \item Differential Calculi on Finite Groups
   \item Metrics
   \item Lagrangian Field Theory and Symplectic Structure
   \item Scalar Field Theory and Spectral of Finite Groups
  \end{enumerate}
  \end{abstract}
  \section{Introduction}
   Finite groups provide a type of simple, nevertheless characterizing enough
   models for {\it noncommutative geometry} (NCG) from both physical and mathematical point of
   views.\\

   Physically, both the problem which geometry would possess a physical
   realization, hence being descended to the real world, and the
   problem which geometry could be abstracted out from the realistic
   recognition of the physical laws are profound and ever being
   asked by generations of physicists who pursue the perfect unity between
   the laws and the languages. Microscopic description of the {\it space-time}, whose picture, after
   the great success of general relativity, was legalized as a differential Lorentzian 4-manifold,
   was considered bit by bit by some outstanding theorists, which
   can be categorized into two lines according to Madore
   \cite{Mad}.\\
   \begin{enumerate}
   \item Modification of the structure of space-time, which can be
   traced back to Dirac \cite{Dir}, Snyder \cite{Sny}, Yang
   \cite{Yang};
   \item Extension of space-time, to which immediately one could recall Kaluza and
   Klein \cite{KK}, Manton \cite{Man} and, appearing ten years
   ago, Connes and Lott \cite{CL}.
   \end{enumerate}
   Along both lines, noncommutative geometry will play a crucial role as
   the linguistic foundation, once if the geometric picture
   deviates from a manifold, being differential or just being topological.
   And finite groups are by far among those which appear in both circumstances
   as the most convenient and manipulable specific models.\\

   Mathematically, the famous {\it Gelfand-Na\v{\i}mark theorem}
   bridges (locally) compact Hausdorff spaces on the geometric bank
   and (unital) commutative $C^\ast$-algebras on the algebraic
   bank \cite{GM}. The philosophy that people learn from this relation is that geometric information
   is able to be encoded in algebraic structure; therefore, the
   generalization of this result to noncommutative regime gives
   birth to noncommutative geometry \cite{NCG}\cite{Mad}.
   To mathematicians, NCG is not only a novel method employing fully the
   power of algebraic tools, whose origin could be even retrieved to Descartes, but also providing a class of completely
   new math objects, hence a completely new scientific scope.
   Within all kinds of approaches towards the implementation of
   the NCG idea, the noncommutative differential geometry on finite
   groups is, in our understanding, the most accessible, however by no means being trivial.\\

   This report is organized in the following way. First in Sect.
   \ref{Sec_A}, mathematical foundation of NCG over finite groups
   is established. Then Section \ref{Sec_B} is dedicated to
   consider metrics in NCG over finite groups. As physical
   applications, lagrangian field theory over finite groups is derived in Section \ref{Sec_C}, together with an induced
   (multi-)symplectic structure, while harmonic analysis of finite
   groups is considered in Sect. \ref{Sec_D}. Some open
   discussions are put in Sect. \ref{Sec_E}.
  \section{Differential Calculi on Finite Groups}
  \label{Sec_A}
  Mathematical framework of NCG for finite sets, especially for finite groups,
  is developed in a series of work \cite{NCGFG},
  and here the formalism mainly follows \cite{Song}.
  \subsection{Group Function}
   Let $G$ be a finite set. $\Alg(G)$ is the algebra of complex
   functions on $G$, whose multiplication is defined pointwisely.
   The basis of $\Alg(G)$ is a collection of delta functions
   \[
    e^g(a)=\dt^g_a, \forall g, a\in G
   \]
   and the algebraic structure of $\Alg(G)$ is characterized by
   \[
    e^g\cdot e^h=e^g\dt^{gh}, \sum_ge^g=1
   \]
   where $1$ is the multiplication unit of $\Alg(G)$.
   A group structure can be implemented on $G$ by extending $\Alg(G)$ to be {\it Hopt
   algebra}, namely for all $f,g\in\Alg(G)$ and $a,b\in G$,
   \[
    M:\Alg(G)\ot \Alg(G)\rightarrow \Alg(G), M(f,g)(a)=f(a)g(a), M(f,g)\equiv f\cdot g\equiv fg
   \]
   \[
    \triangle: \Alg(G)\rightarrow \Alg(G)\ot \Alg(G),
    \triangle f(a,b)=f(m(a,b)), m:G\ot G\rightarrow G, (a,b)\mapsto
    m(a,b)=:ab
   \]
   \[
    1^\pr: \Alg(G)\rightarrow \complex, f\mapsto f(e), e\in G
   \]
   \[
    s:\Alg(G)\rightarrow \Alg(G),
    s(f)(a)=f(\i(a)), \i\in ISO(G)
   \]
   in which $e$ is a specified element in $G$ and $ISO(G)$ is the isomorphism group of $G$ as a set. It
   is easy to verify this follow claim.
   \begin{proposition} \cite{Song}
   $G$ is a group under $(m, e,\i)$ iff $\Alg(G)$ is a Hopf
   algebra under $(M, \triangle, 1^\pr, s)$, i.e. the following
   consistent conditions are satisfied
   \eqa
    \label{E1}
    (\triangle\ot id)\triangle=(id\ot\triangle)\triangle  \\
    \label{E2}
    (1^\pr\ot id)\triangle=id=(id\ot1^\pr)\triangle\\
    \label{E3}
    M(id\ot s)\triangle=1^\pr=M(s\ot id)\triangle
   \ena
   \end{proposition}
   \begin{remark}
   Eqs.(\ref{E1})(\ref{E2})(\ref{E3}) as properties of Hopf algebra correspond to
   associativity, existence of unit and existence of inverse map,
   which are nothing but group axioms. Therefore, the group structure on $G$ can be encoded into the
   structure of Hopt algebra.
   \end{remark}
   Left and right actions on $G$ are set isomorphisms of $G$
   defined as
   \[
    L_g(g^\pr):=gg^\pr, R_gg^\pr:=g^\pr g, \forall g, g^\pr \in G
   \]
   which can be {\it pulled back} to be $L^\ast_g, R^\ast_g$
   acting on $\Alg(G)$ in the canonical way.\\

   Note here we introduce the notation $\bar{g}:=\i(g)$ and will use below.
  \subsection{First Order Differential Forms}
   \begin{definition}
   A first order differential calculus is a pair $(M,d)$, in which
   $M$ is a {\it bi-module} over $\Alg(G)$ and $d$ is a linear homomorphism called {\it
   differential} $d: \Alg(G)\rightarrow M$ satisfying Leibnitz
   rule
   \eq\label{Leib}
    d(ff^\pr)=d(f)f^\pr +fd(f^\pr), \forall f, f^\pr\in \Alg(G)
   \en
   The action of $1$ on $M$ is required to be identity map.
   The elements in $M$ are called first order differential forms.
   \end{definition}
   Universal first order differential is defined by
   $\Omega^1_u(\Alg(G)):=\Alg(G)\ot\Alg(G)/\sim$, and $d_uf=1\otimes f-f\otimes 1$, $d_uf(a,b)=f(b)-f(a)$
   where the equivalent relation is given by $F(a,a)\sim 0, \forall
   F\in \Alg(G)\ot\Alg(G)$.
   \begin{proposition}(Universality of $(\Omega^1_u, d_u)$)
   \cite{DM}\\
   There exists a linear homomorphism ${\mathcal R}$ from
   $\Omega^1_u$ to any $M$, which is referred as {\it reduction} map, such that the follow diagram commutes
   \[
     \CDalign{\Alg(G)&\CDto     &(\Omega^1_u,d)\cr
                     &\searrow  &\CDdown\CDrlabel{{\mathcal R}}\cr
                     &          &(M,d)}
   \]
   \end{proposition}
   \begin{remark}
   The structure of $(M,d)$ can be descended from the universal forms by reduction, namely
   a collection of annihilation relation $e^gd(e^h)=0={\mathcal R}(e^gd_u(e^h))$ for some pairs $(g,h)$.
   \end{remark}
   Left actions are pulled back onto $\Omega^1_u$ by
   \[
    (L_g^\ast\omega)(a,b)=\omega(ga,gb), \forall g, a, b\in G,
    \omega\in\Omega^1_u
   \]
   To generic $M$, left actions are defined by the following rules
   \[
    L^\ast_g(fv)=L^\ast_g(f)L^\ast_g(v), L^\ast_g d=dL^\ast_g
   \]
   for all $g\in G$, $f\in \Alg(G)$, $v\in M$. Similar for inducing right
   actions onto $M$.
   {\it Left-invariant forms in $\Omega^1_u$} are defined by
   \[
    (L_g^\ast\omega)(a,b)=\omega(a,b)
   \]
   It is obviously that all left-invariant forms form a linear
   subspace in $\Omega^1_u$.
   \begin{proposition} \cite{Cas}
    The space of left-invariant forms is $(|G|-1)$-dimensional,
    whose basis is
    \[
     \chi^g=(s\ot id)\triangle(e^g), g\neq e
    \]
    This basis is also a module basis.
   \end{proposition}
   Similarly, right-invariant forms are defined as
   \[
    \eta^g=(id\ot s)\triangle(e^{\i(g)})
   \]
   Below just left-invariant forms will be used.
   \begin{proposition} \cite{Cas}\\
    i) $d_uf=\sum_{g\in G^\pr}\partial_gf\chi^g$,
    where $\partial_g=R^\ast_g-id$ and
    $G^\pr:=G\backslash\{e\}$;\\
    ii) $Eq.(\ref{Leib})\Leftrightarrow \chi^gf=R^\ast_g(f)\chi^g$.
   \end{proposition}
   \begin{remark}(Derivatives on $\Alg(G)$)\\
   It is easy to see that $Der(\Alg(G))$ is trivial. The formal derivatives $\partial_g$ fulfill a {\it deformed Leibnitz
   rule}
   \eqa\nn
   \partial_g(f_1f_2)&=&(\partial_gf_1)f_2+f_1(\partial_gf_2)+(\partial_gf_1)(\partial_gf_2)\\
   \label{Leib0}
   &=&(\partial_gf_1)(\lambda R^\ast_gf_2+(1-\lambda)f_2)+(\lambda f_1+(1-\lambda)R^\ast_gf_1)\partial_gf_2
   \ena
   for any interpolation $\lambda$; however,
   Eqs.(\ref{Leib})(\ref{Leib0}) are compatible iff $\lambda=1$.
   Moreover, even if deformed Leibnitz rule Eq.(\ref{Leib0}) is adopted, $\{\partial_g\}$ is not able
   to be extended to be a left $\Alg(G)$-module, which is the
   well-known property of $Der(C^\infty(V))$ of a differential manifold
   $V$.
   \end{remark}
   A reduction is left-invariant, if there is a subset $G^{\pr\pr}\subset G^\pr$ such that
   ${\mathcal R}(\chi^g)=0$, for all $g\in G^\pr\backslash G^{\pr\pr}$. We will not distinguish left
   invariant basis in $M$ and in universal forms below.
  \subsection{High Order Forms}
   The space of universal p-forms is given by
   $\Omega^p_u(\Alg(G))=\bigotimes^{(p+1)}\Alg(G)/\sim$, where the equivalent relation is that $\omega(a_0,
   a_1,...,a_p)\sim 0$ whenever $a_i=a_{i+1}$ for one
   $i\in\{0,1,2,...,p-1\}$, and the differential is
   $d_u:\Omega^{p-1}_u\rightarrow \Omega^p_u$,
   \eq\label{d}
    (d_u\omega_{p-1})(a_0,a_1,...,a_p)=\sum_{k=0}^p(-)^k\omega_{p-1}(a_0,a_1,...,\hat{a}_k,...,a_p)
   \en
   in which
   $\omega_{p-1}(a_0,a_i,...,\hat{a}_k,...,a_p)=\omega_{p-1}(a_0,a_i,...,a_{k-1},a_{k+1},...,a_p)$.\\

   Universal differential algebra is $\Omega^\ast_u(\Alg(G))=\bigoplus_p\Omega^p_u(\Alg(G))$ where
   $\Omega^0_u(\Alg(G)):=\Alg(G)$, together with the differential $d$.
   A direct calculation shows that
   \begin{proposition}
    Eq.(\ref{d})$\Rightarrow$
    \[
     d\circ d=0
    \]
    \[
     d(\omega^{(p)}\omega^\pr)=d(\omega^{(p)})\omega^\pr +(-)^p\omega^{(p)}d(\omega^\pr)
    \]
    in which $\omega^{(p)}\in\Omega^p_u,
    \omega^\pr\in\Omega^\ast_u$.
   \end{proposition}
   \begin{proposition}
   The cohomology $(\Omega^\ast_u(G),d)$ is trivial.
   \end{proposition}
   \prf
   In fact, consider $\omega_p\in Z^p(\Alg(G))$, i.e.
   $d\omega_p=0$, define $\eta_{p-1}\in\Omega^{p-1}_u(\Alg(G))$ by
   \[
    \eta_{p-1}(a_1, a_2,...,a_p)=\omega_p(e, a_1,a_2,...,a_p)
   \]
   One can check then that $d\eta_{p-1}=\omega_p$.
   \endprf
   \begin{remark}
   Non-trivial topology will emerge if reduction is extended to $\Omega^\ast(\Alg(G))$
   by requiring $d$ to be nilpotent and graded Leibnitz still.
   The resulting quotient algebra is a {\it differential calculus} on
   $\Alg(G)$, which is denoted as $\Omega^\ast(\Alg(G))$. In this
   paper, only the extensions of left-invariant reductions will be
   considered.
   \end{remark}
  \section{Metrics and Covariant Reductions}
  \label{Sec_B}
   \begin{definition}
   An involution $\dag$ on $\Omega^\ast(\Alg(G))$ is defined as
   \[
    f^\dag(g)=\overline{f(g)},
    (d\omega^{(p)})^\dag=(-)^pd(\omega^{(p)\dag}),
    (\omega\omega^\prime)^\dag=\omega^{\prime\dag}\omega^\dag
   \]
   $\forall f\in \Alg(G)$, $\omega^{(p)}\in\Omega^p$,
   $\omega,\omega^\prime\in\Omega^\ast$.
   \end{definition}
   \begin{lemma}
   i) For any $g\in G^{\pr\pr}$,
   $(\chi^g)^\dag=-\chi^{\bar{g}}$;\\
   ii) $X^\dag=X \Leftrightarrow
   R^\ast_g(\bar{X}_{\bar{g}})+X_g=0$;\\
   iii) If $X\in\Omega^1(G)$ is of real coefficients, to which we refer as {\it real}, and exact, then
   \eq\label{her}
     X^\dag=X.
   \en
   \end{lemma}
   Therefore, a reduction is compatible with involution unless
   ${\mathcal R}(\chi^g)=0\Leftrightarrow {\mathcal R}(\chi^{\bar{g}})=0$.
   \begin{definition}
   A metric on $G$ is given by
   $\xi:\Omega^1\ot_{\Alg(G)}\Omega^1\rightarrow \Alg(G)$, if
   $\forall X,Y,Z\in \Omega^1, a\in
    \Alg(G)$, $\xi$ satisfies
    \eqa
     \xi(X+Y,Z)=\xi(X,Z)+\xi(Y,Z)&,&
     \xi(X,Y+Z)=\xi(X,Y)+\xi(X,Z)\\
     \xi(X,a Y)&=&a \xi(X,Y)\\
     \label{real}
     \xi(X^\dag,X)&&\mbox { is real.}
    \ena
   \end{definition}
   \begin{remark}
   i) $\xi$ can be characterized by a set of functions
   $\xi^{gh}$; in fact, for $\forall X,Y \in \Omega^1$
   \[
    \xi(X,Y)=\xi(X_g\chi^g,Y_h\chi^h)=\xi(\chi^gR_{\bar{g}}X_g,Y_h\chi^h)
     =\xi(\chi^g,(R_{\bar{g}}X_g)Y_h\chi^h)
   \]
   \[
     =(R_{\bar{g}}X_g)Y_h\xi(\chi^g,\chi^h)
     =:(R_{\bar{g}}X_g)Y_h\xi^{gh}
   \]
   ii) Similarly, $\xi(Y,X)=X_g(R_{\bar{h}}Y_h)\xi^{hg}$;
   therefore, $\xi(X,Y)\neq\xi(Y,X)$ usually.
   \end{remark}
   A {\it Hermitian structure} on $G$ can be identified as
   $\xi(X^\dag,Y)$.
   \begin{theorem} (Symmetries of metric)\\
   \label{th}
   i) $\xi^{gh}$ are real functions;\\
   ii) $\xi^{gh}=\xi^{\bar{h}\bar{g}}$.
   \end{theorem}
   \prf
   \[
    \xi(X^\dag,X)=\xi((-\chi^{\bar{g}}),X_{g}^\dag
    X_h\chi^h)=-X_{g}^\dag X_h\xi^{{\bar{g}}h}
   \]
   Let $X$ be real, then Eq.(\ref{real}) implies that $\xi^{{\bar{g}}h}$ is
   real.
   So $\xi(X,Y),\xi(X^\dag,Y)$ are real, if $X,Y$
   are real 1-forms.
   Let $X$ be exact, and it can be decomposed into two real exact
   1-forms as $X=X^R+iX^I$, with $X^\dag=X^R-iX^I$ followed from
   Eq.(\ref{her})
   \[
    \xi(X^\dag,X)=\xi(X^R-iX^I,X^R+iX^I)
   \]
   \[
    =\xi(X^R,X^R)+\xi(X^I,X^I)+i(\xi(X^R,X^I)-\xi(X^I,X^R))
   \]
   Therefore, $\xi(X^R,X^I)-\xi(X^I,X^R)$ has to vanish.\\
   Let $X^R=\pt_g f^R\chi^g,X^I=\pt_g f^I\chi^g$, then we get
   \[
    (\pt_g f^I\pt_h f^R-\pt_g f^R\pt_h f^I)\xi^{{\bar{g}}h}=0
   \]
   which implies $\xi^{{\bar{g}}h}=\xi^{{\bar{h}}g}$.
   \endprf
   \begin{remark} (Consistency)\\
   Now three algebraic structures, reduction, involution and
   metric, are specified to $G$.
   For each reduction, there is a constrain on metric which singles out
   some metrics compatible with this reduction, together with involution, i.e.
   ${\mathcal R}(\chi^g)=0\Rightarrow \xi(\chi^g,\chi^h)=0,\xi(\chi^h,\chi^g)=0,
   \forall h\in G$; following theorem \ref{th}, there we have that
   $\xi(\chi^{\bar{g}},\chi^h)=0,\xi(\chi^h,\chi^{\bar{g}})=0,\forall
   h\in G$.
   On the other hand, if $\exists g, s.t.\forall h,
   \xi^{gh}=0=\xi^{hg}$,
   we can induce a reduction which set $\chi^g=0$; also following
   the theorem above, $\xi^{\bar{g}h}=\xi^{h\bar{g}}=0$, implying
   that $\chi^{\bar{g}}=0$.
   \end{remark}
   \begin{definition}
   A reduction, involution and a metric are compatible, if
   \[
    \chi^g=0\Rightarrow \chi^{\bar{g}}=0,
    \xi^{gh}=\xi^{hg}=\xi^{\bar{g}h}=\xi^{h\bar{g}}=0
   \]
   \end{definition}
   Let $\sigma:G\rightarrow G$ be a bijection, the metric/Hermitian structure
   is said to be {\it $\sigma$-covariant} if
   \[
    \sigma^\ast(\xi(X^\dag,
    Y))=\xi(\sigma^\ast(X^\dag),\sigma^\ast(Y)), \forall X,
    Y\in\Omega^1,
   \]
   which implies that
   \eq
   \label{metric}
    \xi^{{\bar{g_1}}g_2}(\sigma(g))=\xi(\sigma^\ast(\chi^{\bar{g_1}}),\sigma^\ast(\chi^{g_2}))(g),
    \forall g\in G
   \en
   Note that such transformations have no infinitesimal forms.
   Below we consider three type of bijections.
   \begin{itemize}
    \item $\sigma$ is a left-translation $L_h$,
    \[ (\ref{metric})\Rightarrow \xi^{{\bar{g_1}}g_2}(hg)=\xi^{{\bar{g_1}}g_2}(g)
    \]
    i.e. $\xi^{{\bar{g_1}}g_2}$s are constant functions on $G$, which can be denoted as
    $\eta^{{\bar{g_1}}g_2}$;
    \item $\sigma$ is a right-translation $R_h$,
    \[
     (\ref{metric})\Rightarrow
     \eta^{{\bar{g_1}}g_2}=\eta^{Ad_h({\bar{g_1}})Ad_h(g_2)};
    \]
    \item $\sigma\in Aut(G)$,
    \eq \label{augorbit}
     (\ref{metric})\Rightarrow \eta^{{\bar{g_1}}g_2}=\eta^{\sigma^{-1}({\bar{g_1}})\sigma^{-1}(g_2)}
    \en
   \end{itemize}
   Define $\gamma^{gh}:=\eta^{{\bar{g}}h}$ which just is a change of symbols, then
   our analysis can be summarized as
   \eq
    \gamma^{gh}=\gamma^{hg},\gamma^{gh}=\gamma^{\sigma(g)\sigma(h)},
   \en
   which implies that if for any $g\in G,\sigma\in Aut(G)$, if a reduction scheme sets ${\mathcal
   R}(\chi^g)=0$,
   then $\chi^{\sigma(g)}$ has to vanish.
   \begin{definition}
    A reduction is covariant, if $G^{\pr\pr}$ contains $Aut(G)$-orbits only.
   \end{definition}
   A covariant metric should be defined upon a covariant
   reduction. In Table \ref{tab1}, covariant reductions for some finite (generated)
   groups are listed.
   \begin{table}[h]
   \caption{Covariant Reductions for Some Finite (Generated) Groups}
   \label{tab1}
   \begin{tabular}{|c|c|c|c|}\hline
    $G$&$Aut(G)$&Num&Covariant Reductions\\\hline\hline
    $Z_2$&$\{e\}$&1&$\chi$\\\hline
    $Z_3$&$Z_2$&1&$\chi^+,\chi^-$\\\hline
    $Z_4$&$Z_2$&3&$\chi^1,\chi^2,\chi^3;\chi^1,\chi^3;\chi^2$\\\hline
    $Z_5$&$Z_4$&1&$\chi^1,\chi^2,\chi^3,\chi^4$\\\hline
    $Z_6$&$Z_2$&7&$\chi^1,\chi^2,\chi^3,\chi^4,\chi^5;\chi^1,\chi^2,\chi^4,\chi^5;$\\\hline
    &&&$\chi^1,\chi^3,\chi^5;\chi^2,\chi^3,\chi^4;\chi^1,\chi^5;\chi^2,\chi^4;\chi^3$\\\hline
    $Z_7$&$Z_6$&1&$\chi^1,\chi^2,\chi^3,\chi^4,\chi^5,\chi^6$\\\hline
    $S_3$&$S_3$&3&$\chi^a,\chi^{a^2},\chi^{\gamma},\chi^{\gamma a},\chi^{\gamma a^2};
     \chi^a,\chi^{a^2};\chi^{\gamma},\chi^{\gamma a},\chi^{\gamma a^2}$\\\hline
    $Z_2\ot Z_2$&$S_3$&1&$\chi^a,\chi^b,\chi^{ab}$\\\hline
    $Z_3\ot Z_3$&$D_4$&1&$\chi^a,\chi^{a^2},\chi^b,\chi^{b^2}$\\\hline
    $Z_3\ot Z$&$Z_2\ot Z_2$&3(nearest)&$\chi^a,\chi^{a^2},\chi^1,\chi^{-1};
     \chi^a,\chi^{a^2};\chi^1,\chi^{-1}$\\\hline
    $(Z)^D_\ot$&$(Z_2)^D\ot_s S_D$&1(nearest)&$\{\chi^{i+},\chi^{i-}|i=1,2,...,D\}$\\\hline
   \end{tabular}\\
   \\
   ``G'': groups; ``Num'': number of covariant reductions.
   \end{table}
  \section{Lagrangian Field Theory}
  \label{Sec_C}
   The following two sections are dedicated to physics of classical field theory on finite
   groups. First integral on $\Alg(G)$ is given by
   \[
    \int_Gf={1\over |G|}\sum_{g\in G}f(g), \forall f\in\Alg(G)
   \]
   {\it Action functional} of a scalar field is $S\in (\Alg(G)\oplus
   \Omega^1(\Alg(G)))^\ast_\real$. Following continuum case, a {\it locality
   principle} is stated as that there exists a {\it lagrangian
   density} $\lag$, such that
   \[
    S(f,\omega)=\int_G\lag(f,\omega),
   \]
   and
   \[
    \frac{\partial S}{\partial f(a)}=\frac{\partial {\cal
    L}(a)}{\partial f}=:\frac{\pt\lag}{\pt f}(a), \frac{\partial S}{\partial \omega_g(a)}=\frac{\partial {\cal
    L}(a)}{\partial \omega_g}=:\frac{\partial {\cal
    L}}{\partial \omega_g}(a)
   \]
   where $\omega\in \Omega^1$ will be take to be $df$
   eventually.\\

   {\it Variation principle} acting on the space of
   $\Omega^0\ot\Omega^1$ by
   \[
    \hat{\dt}\lag(a)=\frac{\partial {\cal
    L}(a)}{\partial f}\hat{\dt}(f(a))+\frac{\partial {\cal
    L}(a)}{\partial \omega_g}\hat{\dt}(\partial_gf(a))
   \]
   \[
    =\frac{\partial \lag(a)}{\partial f}\hat{\dt}(f(a))+\pt_g[R^\ast_{\bar{g}}(\frac{\partial {\cal
    L}}{\partial \omega_g})\widehat{\dt f}](a)-(\pt_gR^\ast_{\bar{g}}\frac{\pt
    \lag}{\pt \omega_g})(a)\hat{\dt}(f(a))
   \]
   \[
    =(\frac{\partial \lag(a)}{\partial f}+(\pt_{\bar{g}}\frac{\pt
    \lag}{\pt \omega_g})(a))\hat{\dt}(f(a))+\pt_g[R^\ast_{\bar{g}}(\frac{\partial {\cal
    L}}{\partial \omega_g})\widehat{\dt f}](a)
   \]
   where $\lambda$ in Eq.(\ref{Leib0}) is chosen to be zero, and
   an identity $R^\ast_g\pt_{\bar{g}}+\pt_g=0$ is used.
   \begin{remark}
   $\hat{\dt}(f(a))$ is differential form upon the space of field variables, $\hat{\dt}((\pt_g
   f)(a))=\hat{\dt}(f(ag)-f(a))=\hat{\dt}(f(ag))-\hat{\dt}(f(a))$.
   Introduce a form-valued function $\widehat{\dt
   f}(a)=\hat{\dt}(f(a))$, then $\hat{\dt}((\pt_g
   f)(a))=(\partial_g \widehat{\dt f})(a)$.
   \end{remark}
   Equation of motion can be read out
   \[
    E(a):=\frac{\pt\lag(a)}{\pt f}+
    (\pt_{\bar{g}}\frac{\pt
    \lag}{\pt\omega_g})(a)=0
   \]
   Twice acting $\hat{\dt}$ on lagrangian gives the result
   \[
    \hat{\dt}^2\lag(a)=\hat{\dt}(E(a))\wedge\hat{\dt}(f(a))
    +\partial_g[\hat{\dt}(R^\ast_{\bar{g}}(\frac{\pt\lag}{\pt \omega_g}))\wedge\widehat{\dt f}](a)=0
   \]
   in which the second term contains a {\it multi-symplectic
   structure} over finite groups.
   \begin{theorem} (Noether theorem on finite groups)\\
   If $\cal{L}$ is invariant under a (vertical) continue transformation $\sigma$ so that
   $\dts f$ is an infinitesimal (not a symbolic differential in above discussion), i.e.
   $\dts{\cal{L}}=\pt_gk^g$, then we have
   \[
    \pt_gJ^g=0,J^g:=(R^\ast_{\bar{g}}\frac{\pt\cal{L}}{\pt\omega_g})\dts f + k^g
   \]
   \end{theorem}
   \prf
   In fact,
   \[
    \dts{\cal{L}}=\pt_gk^g=\frac{\pt\cal{L}}{\pt f}\dts f
     + \frac{\pt\cal{L}}{\pt\omega_g}\dts\pt_g f
   \]
   \[
    =(\pt_g(R^\ast_{\bar{g}}\frac{\pt\cal{L}}{\pt\omega_g}))\dts f
     + (R^\ast_gR^\ast_{\bar{g}}\frac{\pt\cal{L}}{\pt\omega_g})\pt_g\dts f
   \]
   \[
    =\pt_g((R^\ast_{\bar{g}}\frac{\pt\cal{L}}{\pt\omega_g})\dts f)
   \]
   \endprf
   \begin{remark}
   The lesson of this theorem is that there exists a subtlety for the relation between
   conservation laws and symmetries, namely space-time symmetry, since being discrete, e.g. hyper-cubic
   symmetry $\intg_2^d\rtimes S_d$ does imply a conservation
   currents as in continuum.
   \end{remark}
  \section{Scalar Field Theory and the Spectral of Finite Groups}
  \label{Sec_D}
   It is a novel fact that the spectral of Laplacian operator on a
   Riemannian manifold, encodes much, if not all, geometric
   information of this manifold \cite{Ma}. In this section, the
   {\it spectral of finite groups} and their relations to representation theory are explored by defining Laplacian on
   these groups following the results of Sect.
   \ref{Sec_A}, \ref{Sec_B}, \ref{Sec_C} and computing its
   eigenvalues.\\

   A {\it linear scalar field} $\phi$ is an element in $\Alg(G)$. Let $G$ be a
   finite group supported with a left-invariant reduction calculus $G^{\pr\pr}$ and a covariant metric $\xi$,
   then a {\it classical scalar field theory} is defined by
   the following action
   \eqa
   \nn
    S(\phi, d\phi)=\int_G(\xi(d\phi^\dag, d\phi)+V(\phi))\\
   \nn
    =-\int_G({\gamma^{gg^\prime}\pt_g\bar{\phi}\pt_{g^\prime}\phi}-V(\phi))\\
   \nn
    =-\int_G({\bar{\phi}(\gamma^{gg^\prime}\pt_{\bar{g}}\pt_{g^\prime})\phi}-V(\phi))
   \ena
   in which $V(\phi)$ is a generic local function of $\phi$.
   {\it Laplacian} $\vartriangle_G$ specified to $(G, G^{\pr\pr}, \xi)$ is defined
   as
   \[
    \vartriangle_G=\gamma^{gg^\prime}\pt_{\bar{g}}\pt_{g^\prime}
   \]
   and Laplacian equation is
   \[
    \vartriangle_G\phi=\lambda^2\phi
   \]
   Eigenvalues $\lambda^2$ form the spectral of $G$.
   Detailed computations are given below for some examples
   appeared in table \ref{tab1}.
   \subsection{Spectral of $\intg_2$}
    Laplacian equation is given by
    \eqa
    \nn
     \pt\pt\phi&=&\mu^2\phi\\
    \nn
     (R^\ast+(\frac{\lambda^2}{2}-1))\phi&=&0\\
    \nn
     \left(
      \begin{array}{cc}
       \frac{\lambda^2}{2}-1&1\\
       1&\frac{\lambda^2}{2}-1
      \end{array}
     \right)
     \left(
      \begin{array}{c}
       \phi(0)\\
       \phi(1)
      \end{array}
     \right)
     &=&0
    \ena
    where two elements in $\intg_2$ are labeled by $0,1$ and
    $1$ as subscription is omitted. Spectral of $\intg_2$ are
    computed to be
    \[
     \lambda^2=0,4
    \]
    which corresponds to trivial and nontrivial representations of
    $\intg_2$ respectively, namely
    \[
     \phi_{\lambda^2=0}=(1,1)/\sqrt{2}
    \]
    \[
     \phi_{\lambda^2=4}=(1,-1)/\sqrt{2}
    \]
   \subsection{Spectral of $\intg_3$}
    Define $\intg_3=\{0,+,-\}$ and the covariant metric on $Z_3$ is
    \[[\gamma^{gg^\pr}]=
     \left(
      \begin{array}{cc}
       \gamma&\gamma^\prime\\
       \gamma^\prime&\gamma
      \end{array}
     \right)
    \]
    where $\gamma, \gamma^\pr$ are free real parameters.
    Laplacian equation reads
    \eqa\nn
     (2\gamma\pt_+\pt_-+\gamma^\prime(\pt_+\pt_++\pt_-\pt_-)-\lambda^2)
     \phi&=&0\\
     \label{Eq_L3}
     ((2\gamma+\gamma^\prime)(\pt_++\pt_-)+\lambda^2)
     \phi&=&0
    \ena
    Introduce symbols
    $\alpha:=2\gamma+\gamma^\prime,\beta:=2\alpha-\lambda^2$, then
    Eq.(\ref{Eq_L3}) takes the form that
    \[
     (\alpha(R^\ast_++R^\ast_-)-\beta)\phi=0
    \]
    Adopt matrix form,
    \[\Phi:=(\phi(0),\phi(+),\phi(-))^T\]
    \[
     \Box:=
      \left(
       \begin{array}{ccc}
        -\beta&\alpha&\alpha\\
        \alpha&-\beta&\alpha\\
        \alpha&\alpha&-\beta
       \end{array}
      \right)
    \]
    Then Laplacian equation becomes
    \[\Box\Phi=0\]
    whose $det(\Box)=-\beta^3+3\alpha^2\beta+2\alpha^3$. Equation $det(\Box)=0$ gives the spectral
    \[\beta=2\alpha,-\alpha,-\alpha
    \]
    \[
      \Leftrightarrow \lambda^2=0,3\alpha,3\alpha
    \]
    Eigenvectors are
    \eqa\nn
     \phi_0=(1,1,1),\\
     \nn
     \phi^{(1)}_{3\alpha}=(1,\omega,\omega^2),\\
     \nn
     \phi^{(2)}_{3\alpha}=(1,\omega^2,\omega),
    \ena
    corresponding to the three irreducible representations of $Z_3$,
    where $\omega:=-{1\over 2}+i {{\sqrt 3}\over 2}$, the cubic root of unity.
   \subsection{Spectral of $\intg_4$}
   $\intg_4=\{0,1,2,3\}$, and reduction is chosen to be
   $\{\chi^1,\chi^2,\chi^3\}$. Then metric with four free parameters
   is
   \[
    [\gamma^{gg^\pr}]=\left(
      \begin{array}{ccc}
       \gamma^1&\gamma^2&\gamma^3\\
       \gamma^2&\gamma^4&\gamma^2\\
       \gamma^3&\gamma^2&\gamma^1
      \end{array}
     \right)
   \]
   together with Laplacian equation is given by
   \[
    [2\gamma^1\partial_3\partial_1+\gamma^4\partial_2^2+2\gamma^2(\partial_2\partial_1+\partial_2\partial_3)+\gamma^3
    (\partial_3^2+\partial_1^2]\phi=\lambda^2\phi
   \]
   \[
    (2(\gamma^1+\gamma^3)(\partial_1+\partial_3)+2(\gamma^4+\gamma^2-\gamma^3)\partial_2+\lambda^2)\phi=0
   \]
   whose spectral are
   \eq\label{Sp_4}
    \lambda^2=0,4(\gamma^1+\gamma^4+2\gamma^2),4(\gamma^1+\gamma^4+2\gamma^2),8(\gamma^1+\gamma^3)
   \en
   \subsection{Spectral of $\intg_2\ot\intg_2$}
   $\intg_2\ot\intg_2=\{0,a,b,ab\}$, with reduction
   $\{\chi^a,\chi^b,\chi^{ab}\}$.
   Then two-parameter metric is
   \[
    [\gamma^{gg^\pr}]=\left(
      \begin{array}{ccc}
       \gamma&\gamma^\prime&\gamma^\pr\\
       \gamma^\prime&\gamma&\gamma^\pr\\
       \gamma^\pr&\gamma^\pr&\gamma
      \end{array}
     \right)
   \]
   Laplacian equation
   \[
    [\gamma(\partial_a^2+\partial_b^2+\partial_{ab}^2)+2\gamma^\pr(\partial_a\partial_b+\partial_b\partial_{ab}+\partial_{ab}
    \partial_a)]\phi=\lambda^2\phi
   \]
   \[
    (\partial_a+\partial_b+\partial_{ab}+\frac{\lambda^2}{2(\gamma+\gamma^\pr)})\phi=0
   \]
   \[
    \frac{\lambda^2}{2(\gamma+\gamma^\pr)}=0,4,4,4
   \]
   in contrasting with the result in Eq.(\ref{Sp_4}).
   \subsection{Spectral of $\intg_5$}
   $\intg_5=\{0,1,2,3,4\}$, and reduction is $\{\chi^1,\chi^2,\chi^3,
   \chi^4\}$.
   \[
    [\gamma^{gg^\pr}]=\left(
      \begin{array}{cccc}
       a&b&b&d\\
       b&a&d&b\\
       b&d&a&b\\
       d&b&b&a
      \end{array}
     \right)
   \]
   Laplacian equation
   \[
    [2a(\partial_1\partial_4+\partial_3\partial_2)+2b(\partial_1\partial_2+\partial_1\partial_3+\partial_2\partial_4
    +\partial_3\partial_4)+d(\partial_1^2+\partial_2^2+\partial_3^2+\partial_4^2)]\phi=\lambda^2\phi
   \]
   \[
    ((2a+2b+d)\sum_{i=1}^4\partial_i+\lambda^2)\phi=0
   \]
   Spectral
   \[
    \frac{\lambda^2}{2a+2b+d}=0,5,5,5,5
   \]
   \subsection{Spectral of $\intg_6$}
   There are two different six-element groups, $\intg_6$ vs $S_3$.
   $\intg_6=\{0,1,2,3,4,5\}$, and reduction is chose to be
   $\{\chi^1,\chi^3,\chi^5\}$
   \[
    [\gamma^{gg^\pr}]=\left(
      \begin{array}{ccc}
       \gamma^1&\gamma&\sigma\\
       \gamma&\gamma^3&\gamma\\
       \sigma&\gamma&\gamma^1
      \end{array}
     \right)
   \]
   Laplacian equation
   \[
    [2\gamma^1\partial_5\partial_1+\gamma^3\partial_3^2+2\gamma(\partial_3\partial_1+\partial_5\partial_3)+\sigma
    (\partial_5^2+\partial_1^2]\phi=\lambda^2\phi
   \]
   \[
    (2(\gamma^1+\gamma+\sigma)(\partial_1+\partial_5)+2(\gamma^3+2\gamma)\partial_3-(2\gamma+\sigma)(\pt_2+\pt_4)+\lambda^2)\phi=0
   \]
   Spectral
   \eq\label{Sp_6}
    \lambda^2=0,3(2\gamma^1+\sigma), 3(2\gamma^1+\sigma),
    4(2\gamma^1+2\sigma+\gamma^3+4\gamma),
    2\gamma^1+4\gamma+4\gamma^3-\sigma,2\gamma^1+4\gamma+4\gamma^3-\sigma
   \en
   which contain two bi-degenerates.
   \subsection{Spectral of $S_3$}
    $S_3:=\{e,a,a^2, \gamma, \gamma a, \gamma a^2\}$, which is the
    most simple non-Abelian group. Differential on $S_3$ is given
    by
    \[
     df=\pt_\gamma f\chi^\gamma+\pt_{\gamma a}f\chi^{\gamma a}+\pt_{\gamma
    a^2}f\chi^{\gamma a^2}
    \]
    and metric is given by
    \[
     \left(\begin{array}{ccc}
      \gamma^{00}&\gamma^{01}&\gamma^{02}\\
      \gamma^{10}&\gamma^{11}&\gamma^{12}\\
      \gamma^{20}&\gamma^{21}&\gamma^{22}
     \end{array}\right)=
     \left(\begin{array}{ccc}
     \eta&\eta^\prime&\eta^\prime\\
     \eta^\prime&\eta&\eta^\prime\\
     \eta^\prime&\eta^\prime&\eta
    \end{array}\right)\]
    where
    $\pt_0:=\pt_\gamma,\pt_1:=\pt_{\gamma a},\pt_2:=\pt_{\gamma
    a^2}$. Laplacian equation over $S_3$ is
    \[
     (\eta(\pt_0\pt_0+\pt_1\pt_1+\pt_2\pt_2)
     +\eta^\prime(\pt_0\pt_1+\pt_0\pt_2+\pt_1\pt_2+\pt_1\pt_0+\pt_2\pt_0+\pt_2\pt_1)-\lambda^2)\phi=0\\
    \]
    \eq\label{Eq_6}
     (3\eta^\prime(\pt_a+\pt_{a^2})-2(\eta+2\eta^\prime)(\pt_0+\pt_1+\pt_2)-\lambda^2)\phi=0\\
    \en
    Let $c:=2(\eta+2\eta^\prime),d:=6\eta+6\eta^\prime-\lambda^2$,
    Eq.(\ref{Eq_6}) is transformed as
    \[
     (3\eta^\prime(R^\ast_a+R^\ast_{a^2})-c(R^\ast_\gamma+R^\ast_{\gamma a}+R^\ast_{\gamma a^2})
     +d)\phi=0
    \]
    Again define matrix formulated Laplacian equation as
    \[
     \Box:=
      \left(\begin{array}{cccccc}
       d&3\eta^\prime&3\eta^\prime&-c&-c&-c\\
       3\eta^\prime&d&3\eta^\prime&-c&-c&-c\\
       3\eta^\prime&3\eta^\prime&d&-c&-c&-c\\
       -c&-c&-c&d&3\eta^\prime&3\eta^\prime\\
       -c&-c&-c&3\eta^\prime&d&3\eta^\prime\\
       -c&-c&-c&3\eta^\prime&3\eta^\prime&d
      \end{array}\right),
     \]
     \[
     \Phi:=
      (\phi(e),\phi(a),\phi(a^2),\phi(\gamma),\phi(\gamma a),\phi(\gamma
      a^2))^T
    \]
    \[
     \Box\Phi=0
    \]
    Then
    \[
     det(\Box)=-9c^2d^4+d^6+108c^2d^3\eta^\prime-486c^2d^2\eta^{\prime 2}
      -54d^4\eta^{\prime 2}+972c^2d\eta^{\prime 3}
    \]
    \[
     +108d^3\eta^{\prime 3}-729c^2\eta^{\prime 4}+729d^2\eta^{\prime 4}
      -2916d\eta^{\prime 5}+2916\eta^{\prime 6}
     \]
     Vanishing of determinant of $\Box$ gives rise to
     \[
      d=3\eta^\prime,3\eta^\prime,3\eta^\prime,3\eta^\prime,-6\eta^\prime-3c,-6\eta^\prime+3c
     \]
     \[
      \Leftrightarrow \lambda^2=6\eta+3\eta^\prime,6\eta+3\eta^\prime,6\eta+3\eta^\prime,
       6\eta+3\eta^\prime,12\eta+24\eta^\prime,0
    \]
    which contain a quadruple-degenerate, contrasting with the
    spectral in Eq.(\ref{Sp_6}).
    Eigenvectors are organized into three inequivalent irreducible
    representations of $S_3$
    \eqa
    \nn
     \Phi_{trivial}&=&1\oplus 1\\
    \nn
     \Phi_{alternative}&=&1\oplus (-1)\\
    \nn
     (\Phi_{\underline{2}})^i_j&=&
      \left(\begin{array}{cc}
       t\oplus 0&0\oplus t\\
       0\oplus \bar{t}&\bar{t}\oplus 0
      \end{array}\right)
    \ena
    in which any function in $\Alg(S_3)$ is written as
    $f=(f(e),f(a),f(a^2),f(\gamma),f(\gamma a),f(\gamma
    a^2))=:f_+\oplus f_-$, with $f_+:=(f(e),f(a),f(a^2)),
    f_-:=(f(\gamma),f(\gamma a),f(\gamma a^2))$,
    and $1:=(1,1,1),t:=(1,\omega,\omega^2)$.
   \section{Discussion}\label{Sec_E}
    In conclusion, after geometric structures, i.e. differential
    calculi and metric, being introduced onto finite groups, field
    theory is able to be defined, as well as harmonic analysis is
    generalized on the category of finite groups. \\

    The topics of high-tensor fields and spinor fields are not touched in this report. In fact,
    Yang-Mills field can be introduced as one-form \cite{NCGFG};
    however, to define fermions on a generic finite group will
    meet obstacles caused by the possible non-Abelian nature of
    groups. For Abelian cases, a formulation of fermions is given
    in \cite{DS}.\\

  {\bf Acknowledgements}\\
    This work was supported by Climb-Up (Pan Deng) Project of
    Department of Science and Technology in China, Chinese
    National Science Foundation and Doctoral Programme Foundation
    of Institution of Higher Education in China.

  
 \end{document}